OPEN

# Spatio-temporal patterns in growing bacterial suspensions

Pratikshya Jena & Shradha Mishra

**We present a theoretical model to explore the dynamics and phase evolution of growing bacterial suspensions. The model described by the hydrodynamic evolution of bacterial density, orientation, and fluid velocity, incorporating birth and death terms to account for colony growth. Starting from a low-density regime, the system undergoes structural and dynamical transitions driven by bacterial proliferation, leading to the emergence of distinct phases: dilute, turbulent, and heterogenous. At low density, bacteria show local ordering, which transitions to clustering and eventually to a turbulent phase with spatiotemporal vortices of varying scales, as observed in dense suspensions. As density increases, orientation becomes increasingly inhomogeneous and random, indicating heterogeneity. Our findings emphasize the critical role of growth in emergence of phases, illustrating how a single system can sequentially transit through different phases over time. We provide a comprehensive understanding of the evolving patterns and behaviors within growing bacterial colonies, which makes our work different from previous studies on dense bacterial suspensions. Since the growth is generic to active system hence the current work bridges theory and experiment, offering insights into the self-organization and emergent phases in active matter systems driven by growth.**

The emerging field of active matter deals with the analysis of the individual and collective properties of self-propelled agents out of equilibrium[1–6]. Among these, the investigation of active suspensions has been an intriguing research subject for the past few years due to their unique behaviour. A collection of swimming bacteria in a solution is a simplest realization of active suspension, which has a substantial impact on the context of active matter. The bacterial bath[7–11] are interesting model to study due to the simplicity in manipulating the experimental control parameters like concentration of bacteria, activity, and swimming behaviour.

Several research endeavors have been conducted on bacterial suspensions and reported a variety of novel nonequilibrium phenomena including emergence of collective motion[12–14], pattern formation[15,16], reduction in viscosity[17], enhancement of diffusivity[18] and energy transfer[19] etc. In particular, in natural bacterial suspensions different characteristics are found on tuning the packing density and activity[20–24]. Systems show transitions from dilute to ordered and swarming phase for low and moderate densities. For larger packing densities, turbulent and jammed phases are observed[21,24]. Apart from the different phases, the occurrence of spatio-temporal patterns resembling the features of active turbulence in systems of micro swimmers is highly unexpected and has been observed by researchers both experimentally and theoretically[20,21,25–28].

In conventional turbulence, the vortices of different sizes are developed when inertial force dominates over the viscous force and need a supply of external energy to maintain it[29–33]. Alternatively, in bacterial suspensions, the energy injection is inherent and turbulence is observed in dense suspensions even at low Raynolds number. This motivated the researchers in recent years to explore the bacterial turbulence in greater details[34–37].

In recent works[36,38] the role of confinement and boundary condition is explored in bacterial suspensions and it was found that confinement stabilizes the turbulence in the dense suspensions. Whereas in the work of Ref.[39] it was shown that in the presence of confinement, a growing bacterial colony shows the wave type patterns in response to complex chemical environment.

The most of the previous studies on bacterial suspensions have focused on the dynamics and patterns at times when growth is suppressed and suspensions are very high densities[20,27,28,34–36]. The effect of growth in bacterial suspensions is barely studied in theoretical models, except few experiments[40–42].

The dynamics of the system during bacterial growth, along with the competition between growth and other relevant time scales, can result in intriguing statistical and dynamical characteristics[43]. In recent experimental work[44], the authors studied the dynamics of growing bacteria at the oil-water interface and discovered the existence of three growth-induced dynamical phases. But a theoretical model to explore the appearance of distinct phases in growing bacterial suspension is still lacking. This motivated us to present our current work.

Department of Physics, Indian Institute of Technology (BHU), Varanasi 221005, India. email: pratikshyajena.rs.phy20@itbhu.ac.in; smishra.phy@iitbhu.ac.in

We develop a phenomenological coarse-grained model for a growing bacterial suspension by writing the hydrodynamic equations of motion for the bacteria local density, orientation and fluid velocity. An additional birth and death term is introduced to take care of growth of colony.

We started with low density and random orientations of bacteria, and over time, local ordering among the bacteria develops as their density increases with growth. As density increases further, the locally ordered state transit to the dynamical vortex state. Further at high densities system becomes heterogeneous with slower dynamics. Thus, based on the structural pattern and dynamics, three distinct phases are identified; namely *dilute*, *turbulent* and *heterogeneous* phase. Notably, the turbulent phase exhibits spatio-temporal vortex structures, which are typically observed in dense bacterial suspensions[21,45,46]. The turbulent phase is the most dynamic phase, shows the presence of vortices of different sizes and enhanced correlations.

## Model

We consider a growing bacterial suspension in *two*-dimensions. The system is modeled using coarse-grained hydrodynamic equations of motion for the local density $\rho(\mathrm{r}, t)$, orientation $\mathrm{P}(\mathrm{r}, t)$ and fluid velocity $\mathrm{v}(\mathrm{r}, t)$ at position r and time $t$. The hydrodynamic equation for the density $\rho(\mathrm{r}, t)$ is:

$$\frac{\partial \rho}{\partial t} + \nabla.(\rho \mathbf{v}) + v_0 \nabla.(\rho \mathbf{P}) = \nabla^2 \rho + g(\rho), \tag{1}$$

the corresponding equation for the local orientation $\mathbf{P}(\mathbf{r}, t)$ is:

$$\begin{aligned} &\frac{\partial \mathbf{P}}{\partial t} \\ &+ (\mathbf{P}.\nabla)\mathbf{P} + \nabla|\mathbf{P}|^2 + (\mathbf{v}.\nabla)\mathbf{P} = D_r\left(\frac{\rho}{\rho_c} - 1\right)\mathbf{P} \\ &- |\mathbf{P}|^2\mathbf{P} - \omega_{\alpha\beta}P_\beta - v_1 v_{\alpha\beta}P_\beta - \sigma_1 \nabla\rho + \nabla^2\mathbf{P}, \end{aligned} \tag{2}$$

and the equation for the fluid velocity $\mathbf{v}(\mathbf{r}, t)$ is

$$\frac{\partial \mathbf{v}}{\partial t} + \mathbf{v}.\nabla\mathbf{v} = \eta \nabla^2 \mathbf{v} - \nabla p + \nabla.\sigma^{total}_{\alpha\beta}. \tag{3}$$

The density equation (1) is a continuity equation with additional term $g(\rho) = \lambda\rho(1 - \frac{\rho}{\rho_0})$ which incorporates birth and death of the bacteria. This type of term was first used by Malthus[47] and consequently, the flocks with birth and death are also known as Malthusian flocks and later the same term is used in other models to incorporate the birth and death in the population dynamics[4,6,48]. We fix the $\rho_0 = 10$, such that mean maximum packing density is fixed to 10. The term $g(\rho)$ is, a density dependent growth term, with an effective growth rate $\lambda_{eff} = \lambda(1 - \rho/\rho_0)$ is already observed in growing bacterial colony[41,43], where for low density growth is fast, due to abundant of food. As the number increases, due to crowding and competition, the growth slows down. The second term $\nabla.(\rho \mathrm{v})$ in Eq. (1) refers the convection with response to the fluid flow and the third term $v_0\nabla.(\rho \mathrm{P})$ is the convection due to the bacteria's own orientation. Here, $v_0$ is the self-propulsion speed of bacteria. The first term on the right hand side is the diffusion term, which tries to homogenies the density of bacteria in the suspension.

The Eq. (2) presents the dynamics of orientation of bacteria and is quite similar to the Toner–Tu equation[49] for flocking with few additional terms due to the coupling of orientation field with background fluid velocity. The two convective non-linearities on the left hand side $(\mathrm{P}.\nabla)\mathrm{P}$ and $\nabla|\mathrm{P}|^2$ are present due to absence of Galilean invariance in these systems[49,50]. The last term on the left hand side is the convective non-linearity due to fluid velocity. The first two terms and the last term $\nabla^2\mathrm{P}$ on the right hand side are the deterministic part of time dependent Ginzburg Landau (TDGL) model[51] for orientable rod type bacteria's with Landau-Ginzburg free energy functional[52]. In addition two terms on the right hand side $\omega_{\alpha\beta}=\frac{1}{2}(\partial_\alpha v_\beta - \partial_\beta v_\alpha)$ and $v_{\alpha\beta} = \frac{1}{2}(\partial_\alpha v_\beta + \partial_\beta v_\alpha)$ represent, how the flow affects the local orientation of bacteria and are called the vorticity and strain-rate tensors respectively. The term, $\nabla\rho$ is the inertial coupling in local orientation of bacteria in the suspension.

The equation for the fluid velocity equation 3 contains the terms present in Navier-Stokes equation with the additional force term due to the active and passive stresses in the suspension. The origin of the term $\nabla.\sigma^{total}_{\alpha\beta}$ is due to the stresses experienced by the fluid including the internal stress due to viscosity and also the external stress by the suspended bacteria. In this case the total stress $\sigma^{total}_{\alpha\beta} = \sigma_{\alpha\beta} + \sigma^a_{\alpha\beta}$, $\sigma_{\alpha\beta}$ and $\sigma^a_{\alpha\beta}$ are the passive and active contributions respectively. The constitutive equation for stress tensor is formulated as:

$$\sigma^{total}_{\alpha\beta} = 2\eta_1 v_{\alpha\beta} + \frac{1}{2}\nu(P_\alpha h_\beta + P_\beta h_\alpha - \frac{d}{2}(P_\gamma h_\gamma \delta_{\alpha\beta})) + \zeta q_{\alpha\beta}. \tag{4}$$

The first term corresponds to the internal stress caused by the viscosity, the second term is the passive stress with the flow coupling coefficient $\nu$ attributable to any suspended passive particles. The h is the field conjugate to the local orientation P. Additionally, the term $\zeta q_{\alpha\beta}$ is the active stresses where $q_{\alpha\beta} = P_\alpha P_\beta - \frac{1}{2}\delta_{\alpha\beta}$. The positive and negative values of $\zeta$ correspond to the ex -tensile and contractile stresses respectively in most bacterial suspension[1]. For $\zeta = 0$, the hydrodynamic coupling is purely passive. The intrinsic length and time scales in the system is set by persistent length $l_0 = \frac{v_0}{D_r} = 0.4$ and $\tau = 1/D_r = 2$ respectively. Later all the lengths and times can be defined in terms of $l_0$ and $\tau$, such that after writing the re-scaled equations, all terms are dimensionless. To solve the system we perform the numerical integration of Eqs. (1), (2) and, equations 1, 3, 2 in SM [SM1] using Euler's scheme with grid size $\Delta x = 0.5 = 1.25 l_0$ and $\Delta t = 0.0025\tau$.

The value of $\lambda = 0.1$ in the simulation. The $\tau$ is 400 times the typical time step. Hence bacteria shows the persistent motion for approximately 400 time steps of motion and reproduces after 2000 time steps of its motion. The reproduction of bacteria is five times slower than its persistent time in a dilute regime (at very low concentration).

The system is initialized with random, isotropic and homogeneous state with initial number density $\rho \in [\langle \rho \rangle \pm 0.05]$, where initial very low mean density $\langle \rho \rangle = 0.1$. The initial state for orientation is random with two components of $\mathrm{P} = (P_x, P_y)$ and scalar field $\psi$ randomly chosen between $\pm 0.1$ and $\pm 0.05$ respectively. The strength of active coupling $\zeta = -2v_0$ and $v_0$ is fixed to 0.2. The numerical simulation is stable for the above values of parameters in the simulation. The critical density $\rho_c$ beyond which the ordering starts to dominate is fixed to 0.5. The coefficient of inertial term $\sigma_1$ is set to 1 for most of the detailed results. To see the effect of inertia later it is varied from 0.1 to 100. For $\sigma_1 \leq 5$ we see the consistent results and larger value of $\sigma_1$ inertia dominates.

The simulation is carried out for different system sizes *viz* $K$ = 256, 600 and 1024 with period boundary conditions in both the directions and for total simulation time $t = 3000 = 1500\tau$. All the real snapshots in the Results section are generated for $K = 256$. The one simulation step is counted when Eqs. (1), (2), and Eqs. 1, 3 in SM [SM1] are updated once for all lattice points. Statistical averaging is performed over 100 independent realisations for better statistics.

## Results

Starting with the random homogeneous initial state and low mean density, we first analyzed the snapshots of the system at different times. Since, the density of bacteria is increasing with time, all the measurements are done with respect to the corresponding mean density $\langle \rho(t) \rangle$ of the bacteria in the system. In Fig. 1a–d, we showcase the snapshots of the local density with the local orientation at mean densities $\langle \rho \rangle$ = 0.12 , 5.2, 7.5 and 10.0 respectively. The color code shows the local density of bacteria $\rho(\mathrm{r}, t)$ and arrows are the magnitude and orientation field $\mathrm{P}(\mathrm{r}, t)$. In Fig. 1a, when the density of bacteria is low $\langle \rho \rangle = 0.1$ the bacteria are sparsely dispersed with almost homogeneous density in the system with negligible local clustering and local orientation. Then with time as density grows, the small local alignment among the bacteria starts to develop as shown in Fig. 1b for $\langle \rho \rangle = 5.2$. As the density grows further vortex type patterns start to develop in the orientation field of particles, as shown in Fig. 1c at $\langle \rho \rangle = 7.5$ and as density increases further system becomes heterogeneous with random orientation as shown in Fig. 1d for $\langle \rho \rangle = 10.0$. The bottom panel snapshots Fig. 1e–h are the zoomed version of the Fig. 1a–d respectively. In Fig. 1a–d, the length of the arrows are the magnitude of the local orientation**P.**whereas in the bottom panel, the length of the arrows are kept the same for better visualisation of the direction of local orientation **P**. Clearly the directions of arrows in bottom panel show the almost random orientation for very small mean density $\langle \rho \rangle = 0.12$ and orientation correlation starts to develop for intermediate density $\langle \rho \rangle = 5.2$ and maximum for $\langle \rho \rangle = 7.5$ and again randomizes for high density $\langle \rho \rangle = 10$. For clarity arrows are drawn with averaging the local $\mathrm{P}(\mathrm{r})$ in a cell size of $12 \times 12$. In SM [SM2] Fig. 1a–d we show the snapshots for local magnitude of fluid velocity $|V|$, where arrows represent the direction of **P**. Figure 1e–h are the zoomed snapshots for better visualization of local structures.

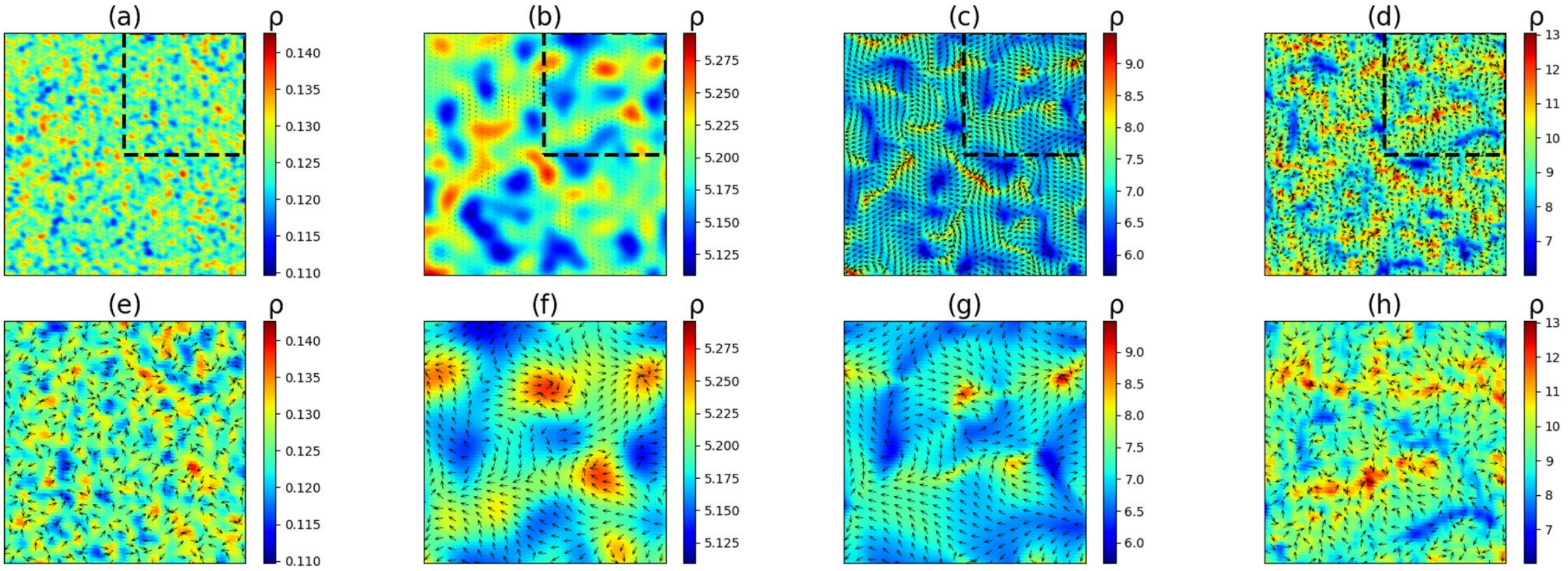

**Fig. 1**. The plots (**a–d**) showcase the real snapshots of bacterial density at different mean densities of bacteria i.e, $\langle \rho \rangle$ = 0.12, 5.2, 7.5, 10 respectively. in the system. The color on the heatmap presents the values of local density at each lattice point. The black arrows present the local bacterial orientation at that lattice point. The length of the arrows are proportional to the magnitude of local P. The plots (**e–h**) illustrate the zoomed portion (square box) of the above snapshots in sequence. For better visualization, we have drawn the arrows by coarse-graining the lattice points in the cell of size $12 \times 12$ and shown them in equal magnitude in (**e–h**). These data are obtained for a single simulation frame.

The above snapshots are observed at four distinct times or at four different mean densities of the system $\langle\rho\rangle$. Next, we focus on the properties of Bacterial suspension with continuous increase of density. Figure 2a–d illustrates the root mean square fluctuations of the local order parameter $P_{rmsf}(\langle\rho\rangle)$, normalized fluid velocity $v_{rmsf}(\langle\rho\rangle)$ , density $\Delta\rho$ and normalized vorticity $\bar{\Omega}_{rmsf}(\langle\rho\rangle)$ vs. $\langle\rho\rangle$ respectively. $\Delta\rho$ is obtained by calculating the fluctuations (variance) in local density by dividing the whole system in boxes of size $4\times4$. The $P_{rmsf}$ , $v_{rmsf}$ and $\bar{\Omega}_{rmsf}$ is obtained by calculating the fluctuations in magnitude of global P and v and $\omega$, where global quantities are obtained by averaging the magnitude of local orientation, fluid velocity and vorticity over all the points in space. The plots suggest that all four fields vary non-monotonically with the mean density $\langle\rho\rangle$. For small mean densities all four fields are small and starts to develop as density increases and shows the maxima at some intermediate density and then further decay at large densities.

Observing such non-monotonic density dependence of root mean square fluctuations, density fluctuations and spatiotemporal snapshots of local density shown in Fig. 1a–h, we identified the three distinct phases in the system named (i) dilute, (ii) turbulent, and (iii) heterogeneous in chronological order. The vertical ticks (blue) in Fig. 2a–d are the mean densities at which the snapshots in Fig. 1a–h are generated. The transition from one phase to other is only qualitative and system simply shows a crossover from one type behavior to other. No sharp phase boundary can be predicted, the same as in previous studies[21,53]. In Fig. 2a–d, the first vertical dashed line is a point where we marked the crossover from dilute to turbulent regime. The velocity fluctuations start increasing from its constant flat value. And again the second dashed line is when the system crosses from turbulent to heterogeneous phase when the local ordering of bacterial orientation starts decreasing from its peak value. On the other hand, it can also defined by looking at the nature of the curves in Fig. 2a–d and density and orientation pattern present in the system.

In the dilute phase, the local density of bacteria is low (Fig. 1a,e), homogeneous pattern of density ($\Delta\rho$ small) and there is almost no ordering (random orientation) and zero $P_{rmsf}(\langle\rho\rangle)$, $v_{rmsf}(\langle\rho\rangle)$ and $\bar{\Omega}_{rmsf}(\langle\rho\rangle)$. In contrast, in the Fig. 1b,f, the local fluctuation in density $\rho$, **P** , **v** and $\bar{\Omega}$ starts to grow showing a crossover from dilute to turbulent phase. On further increment of density, in the turbulent phase all the quantities in Fig. 2a–d attain a maxima and density inhomogeneity is maximum (high $\Delta\rho$) as shown in Fig. 1c,g. Later, we will also observe that in this phase swirling structures develop in the system, and finally, the $\Delta\rho$, the $P_{rmsf}(\langle\rho\rangle)$, $v_{rmsf}(\langle\rho\rangle)$ and $\bar{\Omega}_{rmsf}(\langle\rho\rangle)$ starts to decreases as system becomes heterogeneous. We named this phase as heterogeneous phase due to the weaker and random dynamics present in the system, which we will discuss later in the paper.

The phases we mentioned above can be compared with the recent experiments on growing bacteria on the oil-water interface[44]. The plots in Fig. 2a–d are obtained for $K=256$. We also investigated the same for system size $K=512$ and the results are consistent. Till now, we have identified the three phases based on the static properties of the suspension using the local density, orientational ordering and velocity fluctuations. Next, we characterize these phases based on the structural and dynamical properties of the suspension.

We calculated the vorticity field $\omega(\mathrm{r})$ at the specific densities. In Fig. 3 we show the real space snapshots of normalised vorticity $\Omega(\mathrm{r})=\omega(\mathrm{r})/\bar{\omega}$, where $\bar{\omega}$ is the average vorticity in the system. The color bar shows the magnitude of $\Omega(\mathrm{r})$. The lines drawn in figures are the streamline of local orientation. The thickness of streamline is the magnitude of local orientation. The arrows on the stream is the direction of orientation of $\mathrm{P}(\mathrm{r})$. In Fig. 3a,b,d, the range of $\Omega$ is small, and the bacterial orientations appear quite random, as shown by the black arrows. This low magnitude of vorticity indicates a low level of swirling motion. Figure 3c clearly depicts formation of vortices in the system (the magnitude of $\Omega(\mathrm{r})$ increased substantially). Figure 3e–h showcase the zoomed picture of above, where the structures are more prominent in the turbulent phase (g). In [SM3] (Fig. 2), we show the snapshot (Fig. 3g) of vorticity in the turbulent phase, where the range of vorticity is zoomed between [-4:4]. We clearly see the large local fluctuations in vorticity in this phase. The of the vorticity field with streamlines in the system starting from low to high mean density ($\langle\rho\rangle$) is given in https://drive.google.com/file/d/1gAGtajbGO0EpYB5m_MXKZXoNaaQITKEE/view?usp=sharing Movie).

The presence of vortices of different sizes in bacterial suspension is one of the key characteristics of bacterial turbulence. Further, we characterize the statistics of the three phases by measuring the probability distribution function of the magnitude of local vorticity $p(\Omega)$. In Fig. 4, we depict the $p(\Omega)$ *vs.* $\Omega$ for dilute, turbulent and heterogeneous phase on semi-log-$y$ scale and the inset shows the same plot in $\log-\log$ scale. The nature of distribution is very different in turbulent phase in comparison to other two phases. The $p(\Omega)$ decay as a power law with $1/\Omega$ for a range of $\Omega$, with an exponential tail at larger $\Omega$. The Power-law distributions have longer tails than exponential distributions, and the distribution doesn't decay as rapidly even large values are still statistically allowed. The power law shows that there is always a finite probability of having vortices with larger $\Omega$. The clear power law makes the turbulent phase distinct from other two phases and it can be easily measured in experiments.

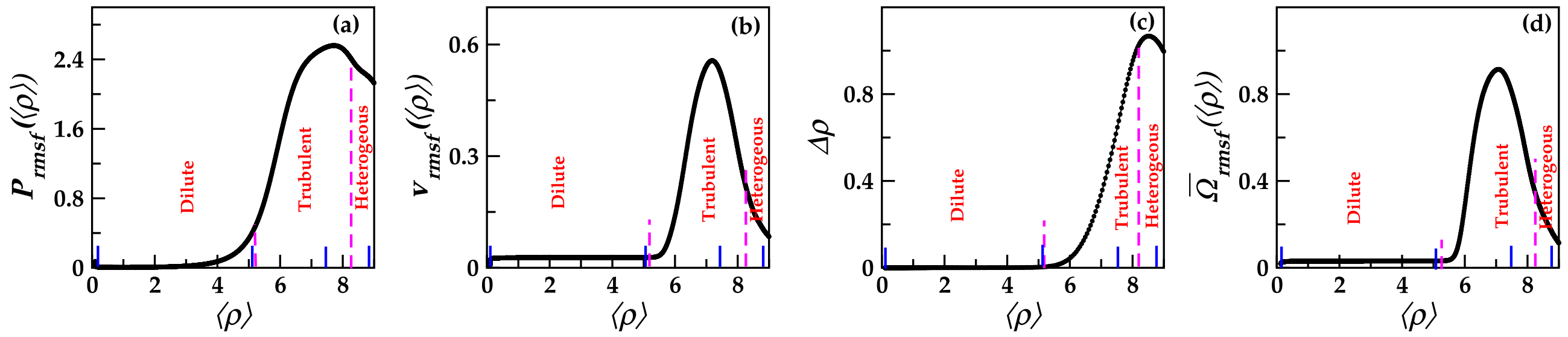

**Fig. 2**. The plot (**a–d**) showcase the fluctuation in root mean square value of global order parameter $P_{rmsf}(\langle\rho\rangle)$, normalized rms of fluid velocity $v_{rmsf}(\langle\rho\rangle)$, $\Delta\rho$ and normalized rms of vorticity $\bar{\Omega}_{rmsf}(\langle\rho\rangle)$ *vs.* mean density $\langle\rho\rangle$ in sequence. The vertical dashed lines (magenta) are showing the boundaries of different phases. The blue short vertical lines in the plots are the points where local density snapshots are shown in Fig. 1a–d. These data are obtained for a single simulation frame.

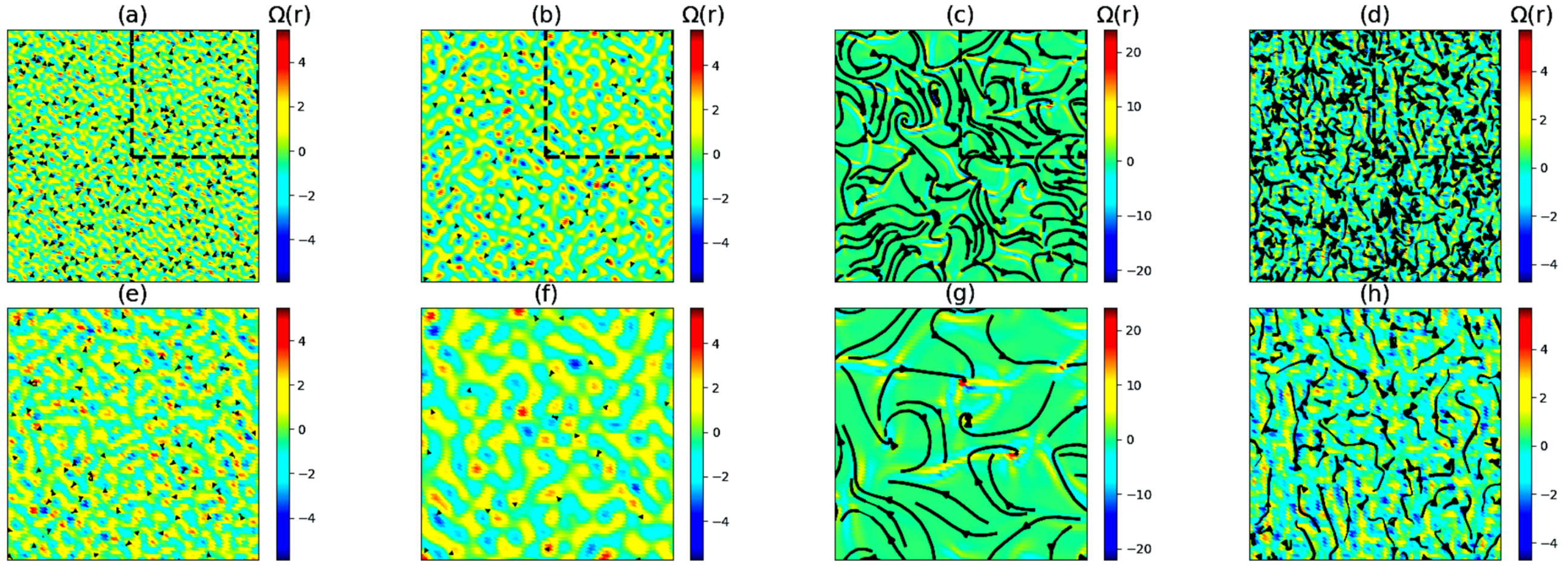

**Fig. 3**. The snapshots (**a–d**) depict the normalized vorticity field $\Omega(\mathrm{r})$ in dilute, cross over to turbulent, turbulent and heterogeneous phases respectively in the system for $\langle\rho\rangle = 0.12, 5.2, 7.5, 10.0$ in sequence. These snapshots are taken at a particular value of global mean density as marked by a blue ticks in the Fig. 2. The plots (**e–h**) illustrate the zoomed portion (square box) of the above snapshots in sequence. The color of the heatmap shows the magnitude of vorticity in all the phases. The lines in plot represent the streamline plot of local orientation $\mathrm{P}(\mathrm{r})$ and the width of lines is the magnitude of local $\mathrm{P}(\mathrm{r})$, the arrows represent the direction of local $\mathrm{P}(\mathrm{r})$. These data are obtained for a single simulation frame.

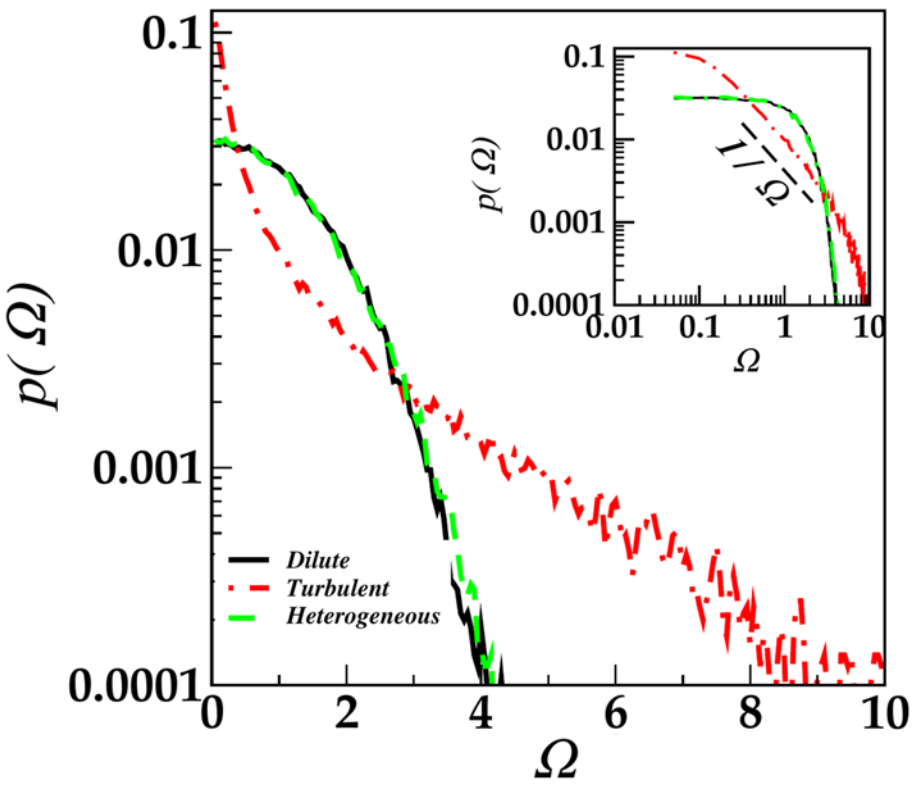

**Fig. 4**. The semi-log-y plot illustrates $p(\Omega)$ of vorticity field $\Omega$ in dilute, turbulent and heterogeneous phases. The legends represent different phases with different line styles. In inset, we plot the log-log plot of the same showing the power-law decay of distribution in the turbulent phase for a range of $\Omega$. These data are obtained for a single simulation frame for $\langle\rho\rangle = 0.12, 7.5, 10$ respectively for dilute, turbulent and heterogeneous phases.

Thus far, our observation is either on the local spatio-temporal pattern or the global order parameters of the system. But, now we aim to show that three phases are indeed the distinct phases and have very different spatial and temporal correlations. To quantify the spatial and temporal correlations in the system as the mean density of bacteria grows with time, we calculate the spatial and temporal correlations of bacterial local density, orientation and fluid velocity. The two-point spatial correlation functions are defined by $C_f(r)$= $\langle \delta f(r_0) \cdot \delta f(r + r_0)\rangle_{r_0}$ where, $f = \rho$, **P** and **v**. The $\delta f$ is the fluctuations from their respective mean values. The angular bracket represents the spherical averaging and average over 100 independent realizations. In Fig. 5a–c, we present the spatial correlation functions of bacterial density $(C_\rho)$ , orientation fields $(C_P)$ and fluid velocity $(C_v)$ in sequence. The color bar shows the mean density of bacteria in the system. The range of density for the different phases are drawn in different symbols. We observe that initially the density correlation is very less in the deep dilute phase and it rises gradually as the density increases in the system showing maximum at the onset of turbulent phase and lowest in heterogeneous phase in Fig. 5a.

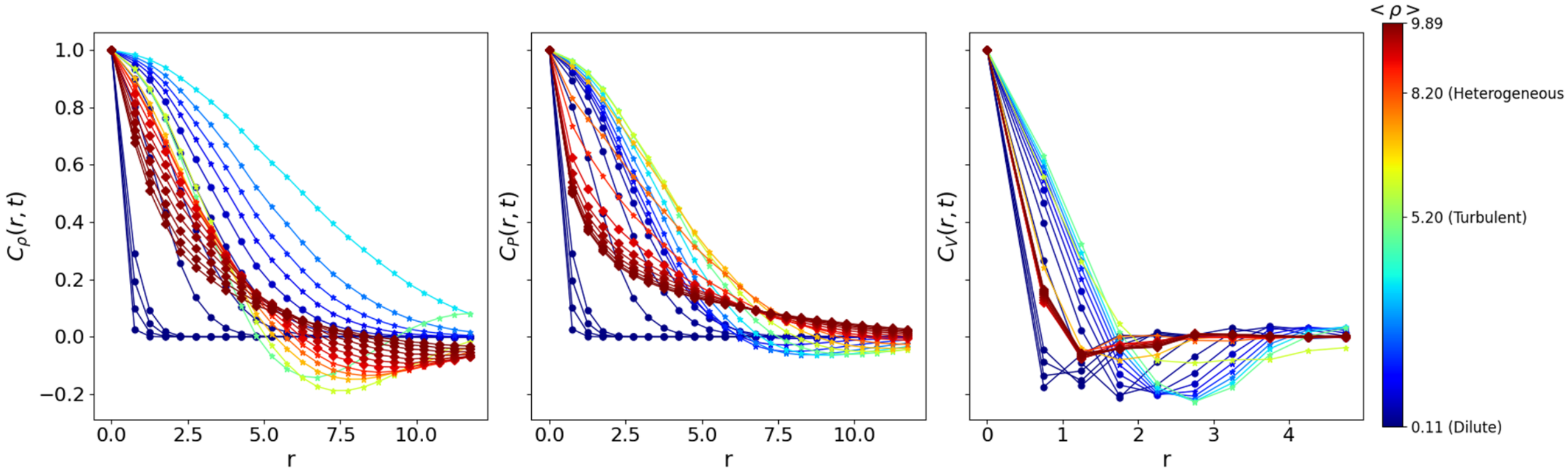

**Fig. 5**. The plots (**a–c**) showcase the spatial correlation functions for bacterial density $C_\rho(r)$, bacterial orientation $C_p(r)$ and $C_v(r)$ respectively in three different regimes i.e, dilute, turbulent and heterogeneous. The correlation functions in dilute, turbulent and heterogeneous phases are drawn by circle (∘), star (∗) and diamond (⋄) symbols respectively. The color bar shows the mean density of the system. And the different phases are mentioned in the color bar.

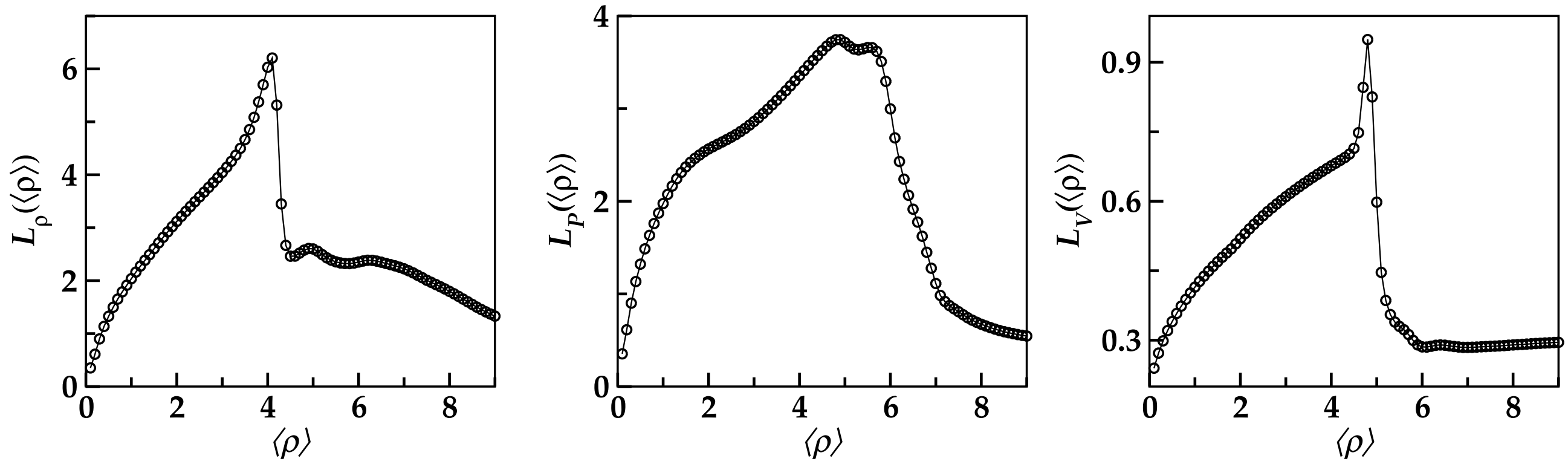

**Fig. 6**. The plots (**a–c**) illustrates the characteristic lengths $L_\rho(\langle\rho\rangle)$, $L_P(\langle\rho\rangle)$ and $L_V(\langle\rho\rangle)$ vs. mean density respectively. The vertical magenta dotted lines show the boundaries of different phases. The results are obtained from system size $1024 \times 1024$. The error bars in these plots are of the size of symbols used.

Starting from very small mean densities, the growth term dominates and population of bacteria grows exponentially with time, that lead to the formation of domains of high density regions gradually and hence at the onset of dilute to turbulent phase all three correlations approaches maximum (Fig. 5a–c) which mimic the faster growth of global fluctuations in density, $P$ and $V$ as seen in Fig. 2a–d. The density correlation starts to decrease at the onset of turbulent and develop negative dips in turbulent phase as can be seen in Fig. 5a–c, which is signature of presence of vortices/antivortices; high/low densities regions as can be seen from the snapshots in Fig. 1c,g. Further the correlations start to deceases as the system transits to the heterogeneous phase at high densities.

We explain the feature of the system as mean density increases in the following manner, starting from the very small mean density, initially for low density growth term $\lambda_{eff}$ dominates over the alignment interaction and hence system observe weaker orientation correlations. With time, effective growth rate decreases with increasing density and it results in competition between active currents (due to self-propulsion speed and fluid velocity) and growth term. Active currents introduce the cohesion among the particles and lead the clustering and local orientational ordering in the system. Hence, two-point spatial orientation and velocity correlation increases. As the mean density grows further with very slow growth rate, growing bacterial colony in a locally ordered region acts like a perturbation and introduces the local defects in the system. With time, it leads to the formation of uncorrelated defects and the system approaches to the disorder high density heterogeneous phase. Almost linear decay of density correlations at large distance suggests the presence of heterogeneous structures in the system as shown in Fig. 5a.

Figure 6a–c demonstrate the characteristic length of density field, orientation field of bacteria and fluid velocity in sequence. We defined the length by the distance by which the the two-point correlation functions decay to 0.5 of its first value. In dilute phase, as we discussed above the bacterial density gradually increases and exhibit a relatively long-range order or higher correlation at the terminal of dilute or the onset of turbulent phase

depicted in Fig. 5a, the correlation length has significantly increases and starts to decrease in the heterogeneous followed by the turbulent phase as clearly observed from Fig. 6a. Further, Fig. 6b,c we show the characteristic length for bacterial orientation $L_P(\langle\rho\rangle)$ and fluid velocity $L_v(\langle\rho\rangle)$ in the system to examine the average size of the structures or patterns generated in the system as the growth of bacteria progresses. From the figures, it is evident that as soon as the system enters the turbulent region, the length $L_P(\langle\rho\rangle)$ and $L_v(\langle\rho\rangle)$ have an increment initially because of the formation of vortices like ordered patterns. At later times, correlations as well as the characteristic length decay. This kind of sharp decay suggests random and chaotic nature of turbulent phase in comparison to other phases.

We further characterize the local correlations among the three fields in the system by calculating the cross-correlations. Figure 3a–c (see in SM [SM4]) shows the cross correlations of density-orientation $C_{\rho P}$, density-fluid velocity $C_{\rho v}$ and bacterial orientation-fluid velocity $C_{Pv}$ respectively. The details of the three correlations are given in SM [SM4]. The three fields are uncorrelated in dilute phase, with zero cross-correlations, whereas as soon as the system transit to the turbulent phase local correlations start to pick non-zero values. The local ordering in the turbulent phase appears at the cost of density inhomogeneity, with high ordered region having lower density and *vice versa* as can be seen by a negative dip in $C_{\rho P}$ in Fig. 3a (see in SM), which is also seen in spatial correlation $C_\rho(r)$ in Fig. 5a. The $C_{\rho v}$ shows a peak in the turbulent phase suggesting the strong local correlation among the bacteria density and fluid velocity. Additionally, the local orientation and fluid velocity ($C_{Pv}$) also show similar trends as density and fluid velocity with a peak in the turbulent phase. As we move towards the heterogeneous phase, $C_{\rho P}$ and $C_{\rho v}$ approaches to zero, because density starts to become homogeneous whereas the local orientation and fluid velocity shows an anti-correlation, due to the local trapping of bacteria.

Till now, we have characterize the properties of all three fields one by one in terms of their statistical properties. In lab designed experiments, in general dynamics of fluid is characterized by measuring the dynamics of a probe particles or going to the frame of tracer particle. Now we use the same approach and look the dynamics of system using Lagrangian approach and characterize the properties of suspension using the dynamics of a probe particle.

In contrast to the Eulerian approach, which focuses on observing the system from a fixed point in space, the Lagrangian method follows individual particles as they move through the system. This approach provides a clearer understanding of the trajectories and local forces experienced by the probe particle, making it particularly useful for studying dynamics in heterogeneous and time-evolving active systems[54].

The dynamics of the probe particle is governed by an over-damped Langevin equation with three contributions to its velocity coming from three factors: (i) local density gradient of bacteria, (ii) fluid velocity and (iii) local orientation of bacteria. The dynamics of probe is affected by the bacterial suspension, but the probe itself is passive and cannot influence the flow of surrounding fluid. Hence, the probe acts like a tracer to study the dynamics of the suspension[55,56]. We started with low density of bacteria in the suspension and placed N =1000 non-interacting probe particles and characterised the speed of probe particle as the density of bacteria in the suspension increases. The details of the governing equation of motion of the probe particles and calculation of speed the of particles is given in SM [SM5]. In the following Fig. 7a, we show the mean speed $U$ of probe particles and $P_{rmsf}$ together vs. $\langle\rho\rangle$ for comparison. As the system goes from a dilute to turbulent phase, the speed increases before decreasing in the heterogeneous phase. The probe particles mirror the system's dynamics. The speed of the probe particles has similar feature as the $P_{rmsf}$. Though, the dynamics of the tracer is influenced by the background flow, bacterial orientation as well density, once the collective dynamics is initiated, the dynamics of bacterial flow is more in comparison to flow of fluid. That's why the tracer particles are more likely to follow the patterns of bacterial orientation rather than the fluid flow. In Fig. 7b, we have shown the mean square displacement (MSD) of the probe particle defined as $\Delta(t)$. In the dilute phase the motion of the particles is ballistic due to their self-propulsion and in the turbulent phase also retains it with a slightly smaller exponent. But, in the heterogeneous phase the motion is arrested due to the trapping of the particles in high density of bacteria. Similar kind of observation is also found in recent experimental works[24,57], where the authors have studied the dynamics of tracer in bacterial suspensions at different densities. In the inset of Fig. 7b, we plot the dynamic exponent of MSD ($\Delta(t) = \frac{1}{N}\sum_i\langle[\boldsymbol{r}_i(t_0+t) - \boldsymbol{r}_i(t_0)]^2\rangle_{t_0}$), $\beta(t)$ defines as $\beta(t) = log_{10}[\frac{\Delta(10t)}{\Delta(t)}]$. The exponent is $\simeq 2$ in the dilute phase and slightly less than that in turbulent phase and finally in the heterogeneous phase we are getting a dip which explains the trapping of the particle at high densities.

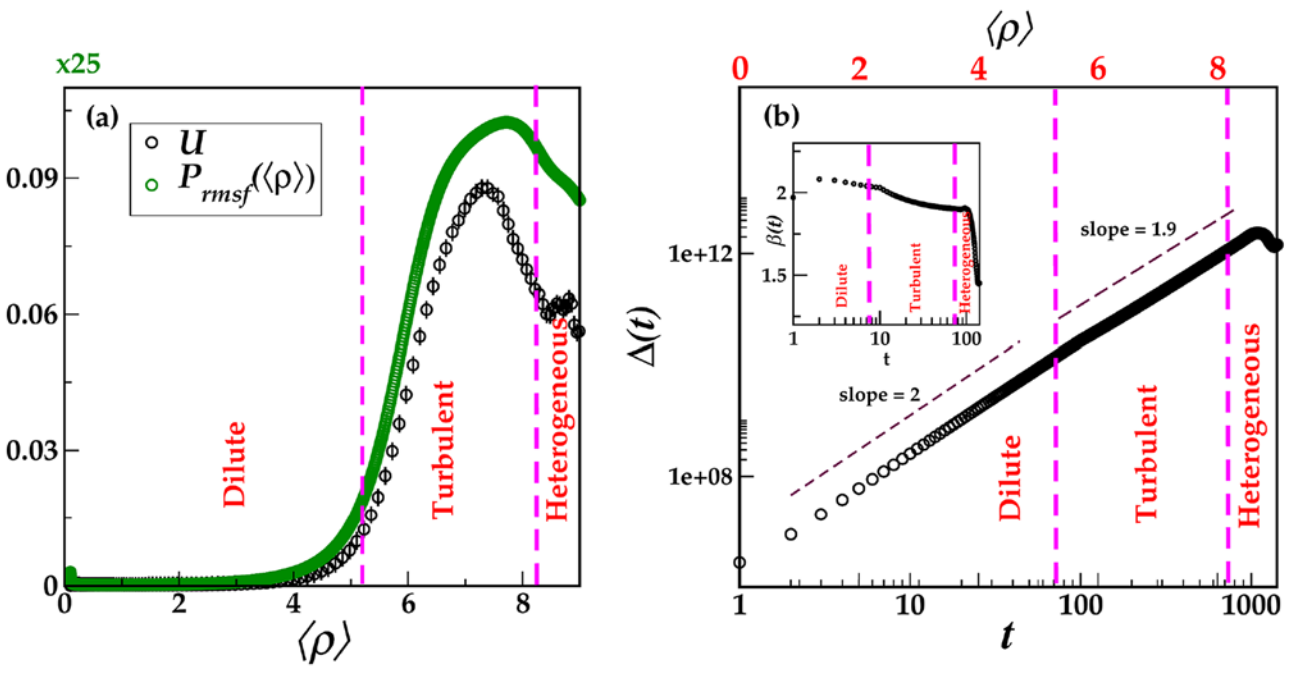

**Fig. 7.** The plot (**a**) showcases the speed ($U$) of the passive probes and $P_{rmsf}(\langle\rho\rangle) vs. \langle\rho\rangle$ as shown by the legends and plot (**b**) presents the MSD ($\Delta(t)$) *vs.*$\langle\rho\rangle$ (upper x-axis) and time (lower x-axis) in sequence with the exponents in different regimes shown by the slopes. The inset plot in (**b**) showcase the plot for $\beta(t) vs. t$ (log scale).

To further quantify the system's energetic, we examine the energy spectrum. So far, it is evident that the turbulent phase is highly dynamic, exhibiting persistent spatio-temporal patterns. The energy spectrum of the fluid velocity is a standard approach to characterise the turbulence in normal fluid. We calculate the energy spectrum of fluid defined by $E(q) = \int < \mathrm{v}(r) \cdot \mathrm{v}(r+R) > \exp(-i\mathrm{q} \cdot \mathrm{R})d\mathrm{R}$. This is employed in order to quantify the spread of energy cascade through the system at different length scales. Figure 8 showcases the same for the different phases in sequence as described above for $\sigma_1 = 1$. The comparison of energy spectrum for other values of $\sigma_1 = 10, 50$ and 100 is shown in SM [SM6]. For the dilute phase, the positive slope at lower $q$ with $E(q) \sim q^2$ indicates increasing energy towards intermediate $q$ and the steep decay at large $q$ shows the quick dissipation of energy at small scales. The $E(q)$ in the turbulent phase, increases at lower $q$ showing a positive slope of value 5 ($E(q) \sim q^5$), which is higher than that of dilute phase. That indicates the rate of increase of energy is more in turbulent phase. The peak of energy density shifts towards lower $q$ value showing the transfer of energy towards large scale from small scales. This is an indicative of inverse energy cascade in $two - dimensional$ active turbulence[58–61]. Here, the source of energy is intrinsic and comes from the active nature of the bacteria in the system. In this phase, we find $E(q)$ shows an intermediate range of $q$ values where $E(q) \sim q^{-2}$. This intermediate $q$ range shows the presence of vortices of different sizes, through which energy is dissipated and further energy dissipation happens sharply as we go for larger $q$ values. Despite the inverse cascade, the energy dissipates at small scales due to viscous dissipation or other mechanism of dissipation showing a negative slope. The range of $q$, where the inverse cascade plays a role can be tuned by tuning the inertia in the system. The bigger range of length scale is found when inertial term is small (small $\sigma_1$) and it is suppressed as we increase $\sigma_1$. For higher value of $\sigma_1$ when system is in turbulent phase as shown in the supplemental Fig. 4 (see in SM [SM7]), the peak of the energy spectrum shifts towards lower value of $q$, representing the accumulation of energy at large scales from small scales is favored by the inertia. Similarly from the plot of the energy spectrum in the heterogeneous phase, we get two different scaling regimes having different slopes. The peaks are at higher $q$ values (smaller scales) in comparison to turbulent phase, the probability of energy transfer from small scales to large scale is less. Different from the turbulent phase, here for a range of intermediate $q$ values energy spectrum $E(q)$ increases with $q$, representing the energy input on relatively larger $q$ values due to high density of bacteria. Further, the peaks of energy spectrum $E(q)$ shifts towards lower $q$ values with increasing value of $\sigma_1$ favoring the accumulation of energy at the larger scales.

Large density inhomogeneities are another interesting characteristics of active systems. To observe the density fluctuations in growing bacteria suspension, we measure the same in the system. We calculate the density fluctuations of bacteria in cells of different sizes. The details of the density fluctuation calculation in given in SM [SM6]. In Fig. 9 we show the plot of density fluctuation $\frac{\bar{\sigma}}{\sqrt{\bar{N}}}$ *vs.* $\bar{N}$, where $\bar{N}$ is the mean number of particles in different cells of the same size. For dilute suspension the density fluctuation is larger for small $\bar{N}$, $\bar{\sigma} \propto \bar{N}$ and attains the equilibrium limit for larger $\bar{N}$, $\bar{\sigma} \propto \sqrt{\bar{N}}$. Hence, at larger distances the density is homogeneous in dilute phase. As system approaches to the turbulent phase, the density fluctuations are much larger at small scales suggesting large density inhomogeneity in the system and further the density fluctuation suppresses at larger $\bar{N}$ ($\bar{\sigma}/\sqrt{\bar{N}}$ decays with $\bar{N}$) showing the local trapping of bacteria at the core of the vortices as can be seen in the snapshot shown in Fig. 1c,g. However, in the heterogeneous phase the maximum density fluctuation is observed at slightly larger N indicating the more correlated heterogenity in density as shown in Fig. 5a and also from Fig. 2c, where $\Delta\rho$ is maximum at the onset of heterogeneous phase. Although the correlation of orientation and velocity fields weakens in the heterogeneous phase, the heterogeneous nature of the system is due to the more inhomogeneous density in the system. We find local regions with very high and lower densities.

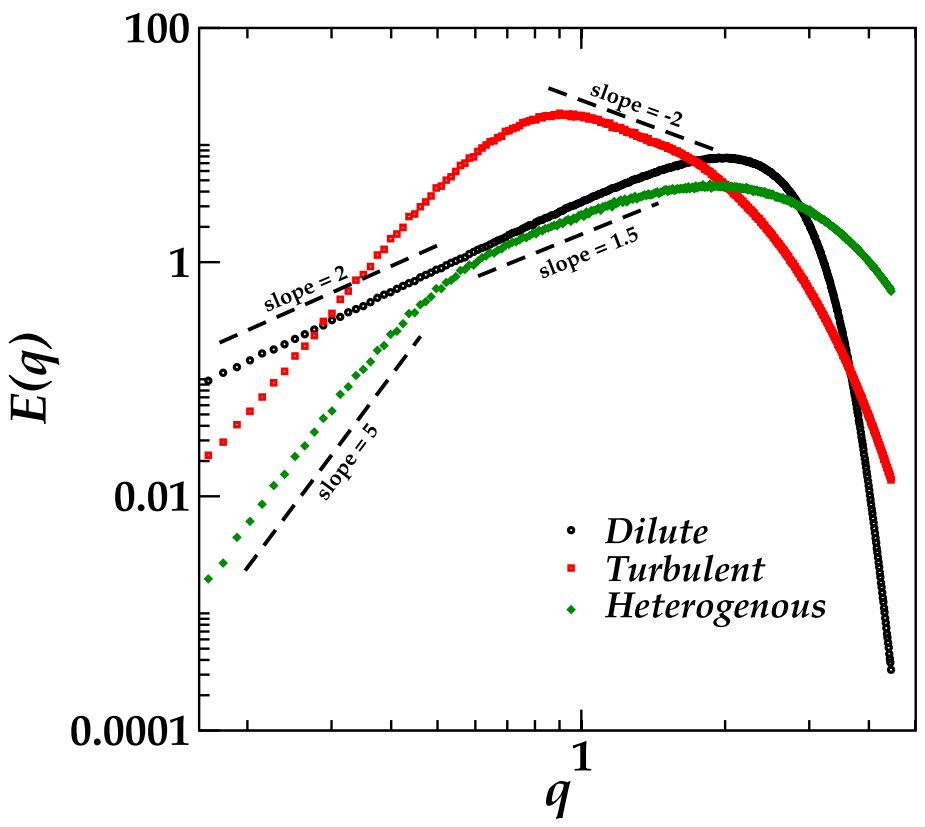

**Fig. 8**. The plot presents the energy spectrum $E(q)$ vs. $q$ in different phases. The $q$ dependence of $E(q)$ is shown by different slopes in different regimes suggesting multiple scaling regimes are present in the system in different phases at $\langle\rho\rangle$ same as mentioned in the caption of Fig. 4. The error bars in this plot are of the size of symbols used.

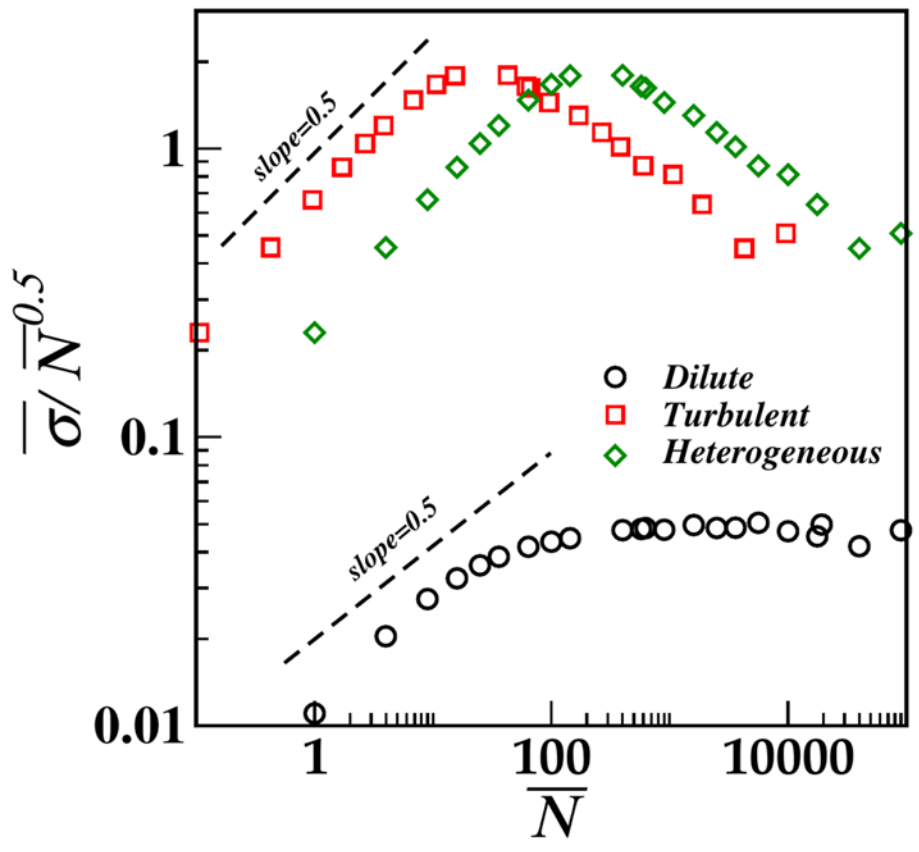

**Fig. 9**. (Color online) The plot depicts the density fluctuation $(\bar{\sigma}/\sqrt{\bar{N}})vs.\bar{N}$ in dilute, turbulent and heterogeneous phases. The large number fluctuations in all phases for small scale are shown by the lines of slopes 0.5. We consider system size K = 600. These data are obtained for a single simulation frame for $\langle\rho\rangle = 0.12, 7.5, 10$ in sequence. $\bar{N}$ is obtained by $\frac{\rho_n}{\rho_{nn}}$. $\rho_n$ is the mean density in each ensembles in the corresponding phase $\rho_{nn}$ is the average of the mean density in the corresponding phase. The details of $\bar{\sigma}/\sqrt{\bar{N}}$ and $\bar{N}$ are given in the SM [SM6].

That makes the density fluctuations larger for the heterogeneous phase. In Fig. 5b (see in SM [SM7]), we show $\bar{\sigma}/\sqrt{\bar{N}}$ *vs.* $\bar{N}$ for large inertia ($\sigma_1 = 10, 50$ and 100) when system is in turbulent phase. As $\sigma_1$ increases, the dip in $\bar{\sigma}/\sqrt{\bar{N}}$ shifts towards larger $\bar{N}$, showing the large scale structures as can be observed in the snapshots in Fig. 3b–d. Additionally, similar information can be analyzed from the energy spectrum $E(q)$ in Fig. 5a (see in SM).

To better understand the results obtained from our model, we present a comparison table below that highlights the differences, similarities, and new insights between our study and previous research.

## Discussion

The active matter system shows interesting individual and collective behaviour. The bacterial suspensions are good *in-vitro* systems to develop controlled experiment to understand the characteristics of active matter. The most of the theoretical and experimental studies on bacterial colony focused on the dense suspensions. But the growth is natural in bacterial colony. Interesting spatio-temporal pattern can emerge, when growth, interaction and activity are competing with each other. Recent experiment by Ref.[44] studies the dynamics and patterns of a growing bacterial suspension on oil-water interface. To replicate the growing bacterial suspension in theory, we proposed a theoretical model to study the statistical and dynamical properties of the growing bacterial suspension by allowing the growth of bacteria in the suspension with time, introduced by a birth and death term in the hydrodynamic equations of motion. The model provides a realistic description of growth dynamics and presents the comprehensive study of the growing bacterial suspension. The key message of our work is that growth can drive the emergence of different phases in bacterial colonies. As growth progresses, the same system transitions through distinct phases over time. A brief summary of previous studies on bacterial suspensions and its comparison to our current study is given in Table 1.

The properties of the suspension is characterised by local density of bacteria, local orientation and fluid velocity. Staring from low density of bacteria, with time the density of bacteria grows and system transits from dilute to the turbulent phase, and at very high density dynamics is suppressed and we obtain the heterogeneous phase. To understand and differentiate the phases that emerge during the growth of bacterial colonies, we employ a set of tools that collectively provide a detailed picture of the system's structural and dynamic properties. Correlation functions reveal how spatial and temporal patterns evolve, capturing transitions between phases. It is observed that local orientation and fluid velocity develop correlation in the turbulent phase with vortex type patterns. Vorticity highlights the emergence of rotational or vortex-like motions in turbulent phase and it has specific spatio-temporal patterns resembling the bacterial turbulence in the previous studies[21,45,46]. Additionally, the rich behaviour of turbulent phase is noticeable by the enhanced dynamics of the probe particle. By examining number fluctuations, we gain insight into deviations from equilibrium behavior in turbulent phase. Additionally, the energy spectrum offers insight into the distribution of kinetic energy across different scales, revealing the dominance of specific modes and patterns in various phases. It shows the presence of dissipation at different length scales due to presence of vortices of different sizes in the turbulent phase and which are absent in other phases. When combined with all the results, it helps bridge the gap between microscopic behaviors and macroscopic patterns, providing valuable insights into the dynamics of growing bacterial colonies.

Further, we found the turbulent nature of the suspension is suppressed, if we increase the effect of inertia and instead wave like patterns are developed[67,68]. The three phases we found in our study and their chacateristics are similar to the different phases obtained in recent experiments on bacterial suspension on oil-water interface[44]. Studying growing bacterial suspensions helps us understand how bacteria grow, interact, and form patterns. The distinct phases we found here can be used to detect the different stages of growth in suspensions. It is useful for industries, such as producing biofuels or biomass production.

| Comparison | Case 1: Dense suspensions (Theory/Exp) | Case 2: Varying concentration constantly (Exp/Simulations) | Our study |
|---|---|---|---|
| Model | Theoretical Studies: Dry active system (Momentum non-conserving)[12,21,46,62]<br>● Focus on hydrodynamic modeling of bacterial turbulence without solvent flow.<br>● Utilizes the Toner–Tu–Swift–Hohenberg (TTSH) equation. Wet active system (momentum conserving)[45,63–65] :<br>● Extends the TTSH equation with momentum-conserving features.<br>● Incorporates the Generalized Navier–Stokes (GNS) equation.<br>● Describes active turbulence in polar momentum-conserving systems. | Varying density in active systems [21,24,53,65,66] :<br>● The experiments, simulations are done by varying the concentrations of the microswimmers constantly in a controlled manner. | Similarities:<br>● Non-monotonic features in local fluctuations in fluid velocity and other characteristics of the turbulent phase.<br>● Evidence of inverse cascade in energy spectra. New findings:<br>● Systematic appearance of different phases as the density grows in the suspension.<br>● Analogous to phases found in 2D growing bacterial suspensions in experiments on oil-water interfaces[44].<br>● Velocity, orientation and density field show non-monotonicity at the onset of turbulence.<br>● Clean power-law decay of local vorticity field observed in the turbulent phase, can be matched in experimental observations.<br>● Comparison of length and time scales in growing suspensions explains the appearance of different phases<br>Differences:<br>● Density field evolves with a birth-and-death term<br>● Unlike the TTSH equation, our model does not include terms responsible for turbulence in the system. |
| Main results | For both wet and dry polar fluid : Active turbulence is found in dense suspensions and density fluctuation is ignored. | In some cases Non monotonicity in velocity field is found with varying concentration. For high densities turbulence is found. | |

**Table 1**. The table showcases a clear comparison among the previous works on dense bacterial suspensions and varying concentration of bacteria with our study of growing bacterial colonies.

In the current study, the suspension evolves in the clean environment, it is interesting to explore the dynamics of the suspension in the presence of complex environment[39].

## Acknowledgements

P.J. and S. M. thank Dr. Sivasurender Chandran for useful discussions. P.J. gratefully acknowledges the DST INSPIRE fellowship for funding this project. The support and the resources provided by PARAM Shivay Facility under the National Supercomputing Mission, Government of India at the Indian Institute of Technology, Varanasi are gratefully acknowledged by all authors. S.M. thanks DST-SERB India, ECR/2017/000659, CRG/2021/006945 and MTR/2021/000438 for financial support. P.J. and S.M. also thank the Centre for Computing and Information Services at IIT (BHU), Varanasi.

## Author contributions

P.J. and S.M. wrote the main manuscript text and P.J. prepared all the figures . All authors reviewed the manuscript.

## Declarations

### Competing interests
The authors declare no competing interests.

### Additional information
**Supplementary Information** The online version contains supplementary material available at https://doi.org/10.1038/s41598-025-13297-5.

**Correspondence** and requests for materials should be addressed to P.J. or S.M.

**Reprints and permissions information** is available at www.nature.com/reprints.